# A fractal classification of the drainage dynamics in thin liquid films

Roumen Tsekov[1] and Elitsa Evstatieva[2]
[1]Department of Physical Chemistry, University of Sofia, 1164 Sofia, Bulgaria
[2]Department of Water Technology, Karlsruhe Research Center, 76021 Karlsruhe, Germany

It is demonstrated that the dynamic structure is very important for the rate of drainage of a thin liquid film and it can be effectively taken into account by a dynamic fractal dimension. It is shown that the latter is a powerful tool for description of the film drainage and classifies all the known results from the literature. The obtained general expression for the thinning rate is a heuristic one and predicts variety of drainage models, which are even difficult to simulate in practice. It is a typical example of a scaling law, which explains the origin of the complicate dependence of the thinning rate on the film radius.

The drainage in thin liquid films is an important process in foams, emulsions and suspensions. Scheludko [1] has adapted to free films the Reynolds formula for the velocity of approach of two parallel rigid discs separated by a liquid slit

$$V = \frac{2h^3(p_\sigma - \Pi)}{3\mu R^2} \qquad (1)$$

where $h$ is the film thickness, $R$ is the film radius, $\mu$ is the liquid viscosity, $p_\sigma$ is the capillary pressure in the meniscus and $\Pi$ is the disjoining pressure. This equation is valid for films with tangentially immobile and plane-parallel surfaces. Deviations from eq 1 due to tangential surface mobility are reported. Since the films are usually stabilized by surfactants, the film drainage causes an interfacial gradient of the surfactant adsorption. Thus, a gradient of the surface tension appears and the corresponding Marangoni effect slows down the interfacial flow. In this case, the theory [2] predicts a thinning rate according to the Reynolds formula from eq 1 but with a reduced viscosity $\eta = \mu(1+aE/D\mu)/(4+aE/D\mu)$, where $a$, $E$ and $D$ are the adsorption length, Gibbs elasticity and diffusion coefficient of the surfactant, respectively. The thickness non-uniformity is another important reason for deviations from eq 1. Some experiments [3] have detected rates of thinning depending much weaker on the film radius as compared to the Reynolds formula. Such a behavior cannot be explained by the surface mobility, the effect of which is not expected to alter the functional dependence of the thinning rate on the film radius. Eq 1 requires strictly symmetrical drainage between two parallel flat surfaces, which is supplied in the Reynolds case by the rigidity of the solid interfaces. In contrast, the shape of free films is determined via the balance of the viscous, capillary and surface forces. Experimental investigations [4] have shown that large films are non-homogeneous in thickness. Due to the film geometry, the liquid in the film center flows slower as compared to that in the film periphery and in this way a characteristic shape is formed, which is known under the name of dimple. The thinning rate of dimpled films is larger than the prediction of the Reynolds law [5].

In the frames of the lubrication approximation, the dynamics of the film drainage is described by the following linearized equation [2]

$$12\eta R^2 \partial_t H = h^3 \Delta P \tag{2}$$

where $H$ is the local film thickness, $h$ is the average film thickness, $P$ is the driving pressure and $\Delta$ is two-dimensional Laplace operator. A relationship between the driving pressure and the local film thickness is further supplied via the normal force balance on the film surfaces

$$P = p - \sigma \Delta H / 2R^2 \tag{3}$$

Here $p$ is the mean driving pressure averaged along the film area and $\sigma$ is the surface tension. Note that the mean pressure $p$ equals to the difference between the capillary pressure in the meniscus $p_\sigma$ and the disjoining pressure $\Pi$. Since $h \ll R$, the films are relatively flat. Hence, in eq 3 the local capillary pressure is proportional to the film curvature and the disjoining pressure contributes by its mean value only. In general, the evolution of the film thickness is governed by eqs 2 and 3 and a set of an initial and four boundary conditions. The latter depend, however, substantially on the film formation. For this reason, there are plenty of different and even contradictive observations of the thinning rate. The scope of the present paper is to classify the laws of thinning by introducing a fractal scale of the dynamics of the film drainage.

Let us begin by the derivation of the Reynolds formula. Supposing the film surfaces are parallel, the local thickness $H$ in the film is everywhere equal to the mean thickness $h$. In this case, eq 2 acquires the form

$$-12\eta R^2 V = h^3 \Delta P \tag{4}$$

where $V = -\partial_t h$ is the mean thinning rate. Integrating twice eq 4 under constrain of zero driving pressure at the firm rim, one yields $P = (3\eta R^2 V / h^3)(1 - x^2)$, where $x = r/R$ is the dimensionless distance from the film center. Combining this result with eq 3 and integrating once again leads to

$$\partial_x H = (pR^2 / \sigma)x - (3\eta R^4 V / 2\sigma h^3)(2 - x^2)x \tag{5}$$

Hence, it follows immediately from the boundary condition $(\partial_x H)_1 = 0$ that the rate of drainage obeys the Reynolds formula. Integrating eq 5 by remembering that $h$ is the mean film thickness, one yields the film profile

$$H = h + (pR^2/12\sigma)(2 - 6x^2 + 3x^4) \qquad (6)$$

As seen, eq 6 describes a dimple with a thicker part in the film center and a thinner ring at the film border. Of course, the film is no more planar and to obtain the range of validity of the Reynolds formula one should require at least a positive thickness in the film thinnest part, i.e. the barrier rim. Using eq 6 one can transform the inequality $H(1) \geq 0$ to

$$R \leq \sqrt{12\sigma h/(p_\sigma - \Pi)} \qquad (7)$$

Therefore, the Reynolds formula is valid only for small films. This is the conclusion of many experiments [3], which clearly show that the Reynolds formula is applicable to sufficiently small films, which are relatively homogeneous in thickness. Note that all the films approach their equilibrium state with the Reynolds rate of thinning, because the inequality above is always fulfilled in the limit $p_\sigma = \Pi$. This is not surprising since the equilibrium films are flat.

The problem now is to describe the drainage of films with radii out of the range of validity of inequality 7. Experimental results [3] indicate an increase of the amplitude of thickness non-homogeneity with increasing film size. This corresponds to growing divergence from the idealized models of planar film profile. Introducing a dimensionless variable $\tau$ via the relation

$$d\tau = (\sigma h^3/24\eta R^4)dt \qquad (8)$$

and substituting eq 3 in eq 2, one yields the following differential equation

$$\Delta\Delta H = -\partial_\tau H \qquad (9)$$

Appling the Laplace transform, this equation changes to

$$\Delta\Delta \tilde{H} = -s\tilde{H} + h_0 \qquad (10)$$

where $\tilde{H}$ is the Laplace image of the local film thickness and the initial profile of the film is set to be flat at thickness $h_0$. If the film is axisymmetric, the solution of eq 10 is a sum of the Bessel functions of zero order

$$\tilde{H} = h_0/s + AJ_0(\sqrt[4]{-s}x) + BI_0(\sqrt[4]{-s}x) + CK_0(\sqrt[4]{-s}x) + DY_0(\sqrt[4]{-s}x) \tag{11}$$

The functions $K_0$ and $Y_0$ are not proper solutions since they posses peculiarity at $x=0$. For this reason, the corresponding coefficients have to be set equal to zero, $C=D=0$. The other two coefficients can be determined by the boundary condition $(\partial_x \tilde{H})_1 = 0$ of minimum at the barrier rim and the definition of the mean thickness to obtain

$$A = \frac{(s\tilde{h} - h_0)\sqrt[4]{-s}}{4sJ_1(\sqrt[4]{-s})} \qquad B = \frac{(s\tilde{h} - h_0)\sqrt[4]{-s}}{4sI_1(\sqrt[4]{-s})}$$

Hence, the Laplace image of the film profile acquires the form

$$\tilde{H} = \frac{h_0}{s} + (s\tilde{h} - h_0)\frac{\sqrt[4]{-s}}{4s}[\frac{J_0(\sqrt[4]{-s}x)}{J_1(\sqrt[4]{-s})} + \frac{I_0(\sqrt[4]{-s}x)}{I_1(\sqrt[4]{-s})}] \tag{12}$$

where the only unknown function is the Laplace image $\tilde{h}$ of the mean thickness. It can be determined from the equilibrium pressure balance at the barrier rim

$$\sigma(\Delta\tilde{H})_1 / 2R^2 = p_\sigma / s \tag{13}$$

For the sake of simplicity, sufficiently thick films are considered here, where the disjoining pressure is negligible as compare to the capillary pressure in the meniscus. Substituting eq 12 in eq 13, one yields the relation

$$s\tilde{h} - h_0 = \frac{8p_\sigma R^2}{\sigma\sqrt{-s}\sqrt[4]{-s}} \frac{J_1(\sqrt[4]{-s})I_1(\sqrt[4]{-s})}{J_1(\sqrt[4]{-s})I_0(\sqrt[4]{-s}) - J_0(\sqrt[4]{-s})I_1(\sqrt[4]{-s})} \tag{14}$$

Thus, the Laplace image of the film profile acquires the form

$$\tilde{H} = \frac{h_0}{s} + \frac{2p_\sigma R^2}{\sigma s\sqrt{-s}} \frac{I_1(\sqrt[4]{-s})J_0(\sqrt[4]{-s}x) + J_1(\sqrt[4]{-s})I_0(\sqrt[4]{-s}x)}{J_1(\sqrt[4]{-s})I_0(\sqrt[4]{-s}) - J_0(\sqrt[4]{-s})I_1(\sqrt[4]{-s})} \tag{15}$$

Unfortunately, it is impossible to invert analytically the Laplace image from eq 15 to obtain the local film thickness. Since the main interested here is the evolution of the mean film thickness, the latter can be expressed from eq 14 in the form

$$h - h_0 = L^{-1}[\frac{8p_\sigma R^2}{\sigma s\sqrt{-s}\sqrt[4]{-s}} \frac{J_1(\sqrt[4]{-s})I_1(\sqrt[4]{-s})}{J_1(\sqrt[4]{-s})I_0(\sqrt[4]{-s}) - J_0(\sqrt[4]{-s})I_1(\sqrt[4]{-s})}] \qquad (16)$$

where the operator $L^{-1}$ indicates the inverse Laplace transformation. A favorite circumstance is that for microscopic films $\tau$ is a very small number. Hence, the Laplace images with large $s$ are mostly important. Expanding the Laplace image in eq 16 for large $s$ allows explicit inversion of the Laplace transformation to obtain

$$h_0 - h = 8(R^2 p_\sigma / \sqrt{2}\sigma)L^{-1}[s^{-7/4}] = 32R^2 p_\sigma \tau^{3/4}/3\sqrt{3}\sigma \qquad (17)$$

This equation can be solved in respect to $\tau$

$$\tau = 9(\sigma/32R^2 p_\sigma)^{4/3}(h_0 - h)^{4/3} \qquad (18)$$

A detailed analysis of eq 18 shows that it implies contribution of the initial film profile on the rate of drainage. This effect is not desired since many experiments show that there are stages in the drainage when the film has already forgotten the initial condition. Fortunately, the dependence of $\tau$ on the film thickness in eq 18 is on power 4/3, which is close to 1. Hence, the following physical approximation of eq 18 is plausible

$$\tau = 12(\sigma/32R^2 p_\sigma)^{4/3}(h_0^{4/3} - h^{4/3}) \qquad (19)$$

which eliminates the disadvantage mentioned above. Substituting eq 19 in eq 8, one yields the velocity of drainage of the films

$$V = \frac{1}{3\eta}\sqrt[3]{\frac{h^8 p_\sigma^4}{2\sigma R^4}} \qquad (20)$$

As seen, the velocity of drainage of large films differs substantially from the Reynolds law. It is inversely proportional to the 4/3 power of the film radius and this weaker dependence on $R$ is in accordance with the experimental observations. Note that for film radii out of the range of validity of inequality 7 the thinning rate $V$ from eq 20 is always larger than the Reynolds estimation.

The rate of drainage in eq 20 is valid for axisymmetric films only. Sometimes, this model does not correspond well to the real film shape, especially in large films formed by quick initial withdrawal of liquid. More complex phenomena [6] take place in such films and the radial symmetry of the thinning process is violated. The asymmetric film drainage is quicker than the axisymmetric one due to the peristaltic pumping effect. Owing to the stochastic nature of the film shape in this case, the description of the film drainage is more complicated and requires a statistical approach. A scaling analysis of eqs 2 and 3 leads to the following equations

$$12\eta R^2 V = h^3 q^2 \delta P \tag{21}$$

$$\delta P = (\sigma h / 2R^2) q^2 \tag{22}$$

where $q$ is the wave number of the thickness perturbations in the film and $\delta P$ is the corresponding changes of the driving pressure. Substituting eq 22 in eq 21, one yields an expression for the characteristic wave number

$$q = (R/h)\sqrt[4]{24\eta V/\sigma} \tag{23}$$

Since the thickness perturbations are generated by the drainage, the wave number $q$ increases with increasing of the thinning rate. Eq 23 demonstrates also that as larger the film is as many the thickness perturbations are. To find out the drainage rate of the films one can employ a scaling law for the driving pressure distribution. Introducing a fractal dimension $\alpha$, the pressure changes can be related to the corresponding wave numbers via the expression

$$\delta P / p = (4/q)^\alpha \tag{24}$$

where the factor 4 corresponds to the average distance between the zeros of the Bessel functions. Eq 24 is a typical example of a power law and implies a geometric correlation between the pressure perturbations. Substituting eq 24 in eq 22 and introducing eq 23 in the result yields the velocity of drainage of the films

$$V = \frac{32\sigma h^4}{3\eta R^4} (\frac{R^2 p}{16\sigma h})^{\frac{4}{2+\alpha}} \tag{25}$$

The law from eq 25 discriminates between different types of drainage by the corresponding dynamic fractal dimension $\alpha$. For instance, if the film possesses two solid interfaces the latter induce strong correlation in both the two directions. In this case $\alpha = 2$ and eq 25 reduces to

$$V = \frac{2h^3(p_\sigma - \Pi)}{3\eta R^2} \tag{26}$$

Eq 26 is in fact the Reynolds law from eq 1. If the drainage is strictly axisymmetric but not confined between solid interfaces, the radial direction is free for relaxation and a dimple occurs. In this case $\alpha = 1$ and eq 25 reduces to

$$V = \frac{1}{3\eta} \sqrt[3]{\frac{h^8(p_\sigma - \Pi)^4}{2\sigma R^4}} \tag{27}$$

Eq 27 is a generalization of eq 20 including the effect of the disjoining pressure. Finally, one can imagine a completely random film broken down to many uncorrelated sub-domains. In this case $\alpha = 0$ and eq 25 acquires the form

$$V = \frac{h^2(p_\sigma - \Pi)^2}{24\eta\sigma} \tag{28}$$

Note, that the rate of drainage of such stochastically corrugated films does not depend on the film radius since any spatial correlation of the dynamics is missing. In a previous paper [5], it was demonstrated that some experiments are well described by the following rate of drainage

$$V = \frac{1}{6\eta} \sqrt[5]{\frac{h^{12}(p_\sigma - \Pi)^8}{4\sigma^3 R^4}} \tag{29}$$

One can easily check that eq 29 corresponds to eq 25 for $\alpha = 1/2$. Hence, the conclusion is that the described films are neither planar nor strictly axisymmetric. The typical shapes of such films possess some kind of regularity [3] but it is far from a strong radial symmetry. For this reason the corresponding fractal dimension is a number between those of axisymmetric and random films.

The present analysis demonstrates that the morphology of the film profiles is very important for the rate of drainage. It can be effectively taken into account by the dynamic fractal dimension $\alpha$, which affects substantially the thinning rate as described in eq 25. As is shown, the latter is a powerful law for description of the film drainage and reproduces all the known results from the literature. Eq 25 is a heuristic one and predicts variety of drainage behaviors, which are even difficult to simulate in practice. It is a typical example of a scaling law, which explains the origin of the complicate dependence of the thinning rate on the film radius. Additional phenomena such as interfacial friction [2] and marginal regeneration [7] can be also included by proper generalization of the effective viscosity $\eta$, which, in this case, will become a thickness dependent quantity.